\documentclass[a4paper,twoside,twocolumn,11pt]{scrartcl}
\usepackage{pic2010_layout}
\usepackage{amsmath}
\usepackage{graphicx,tabularx}
\usepackage{url}
\begin{document}

\renewcommand{\refname}{Selected References}

\def\bentarrow{\raisebox{5pt}{\rlap{$\vert$}}\hspace*{-2.17pt}\to}
\def\nj{n_\text{jets}}
\def\meff{m_\text{eff}}

\newcommand{\etal}{\textit{et al. }}
\newcommand{\ie}{{\sl i.e. }}
\newcommand{\eg}{{\sl e.g. }}
\newcommand{\met}{\slashchar{p}_T}

\def\slashchar#1{\setbox0=\hbox{$#1$}           
   \dimen0=\wd0                                 
   \setbox1=\hbox{/} \dimen1=\wd1               
   \ifdim\dimen0>\dimen1                        
      \rlap{\hbox to \dimen0{\hfil/\hfil}}      
      #1                                        
   \else                                        
      \rlap{\hbox to \dimen1{\hfil$#1$\hfil}}   
      /                                         
   \fi}

\setlength{\floatsep}{0pt}
\setcounter{topnumber}{6}
\setcounter{bottomnumber}{6}
\setcounter{totalnumber}{10}
\renewcommand{\topfraction}{1.0}
\renewcommand{\bottomfraction}{1.0}
\renewcommand{\textfraction}{0.0}

\newcommand{\mytitle}{New physics searches for the LHC}
\newcommand{\myname}{Tilman Plehn}
\newcommand{\myaddress}{Institut f\"ur Theoretische Physik, Universit\"at Heidelberg, Germany}

\twocolumn[
  \begin{@twocolumnfalse}
    \PICtitle{\mytitle}{\myname}{\myaddress}
    \vspace*{-12pt}
    \begin{abstract}
      \noindent
      \textbf{\Large Abstract}
      \vspace*{12pt} \par\noindent\normalsize Data taking at the LHC
      is the beginning of a new era in particle physics which will
      lead us towards understanding the completion of the Standard
      Model at and beyond the TeV scale. I discuss different
      approaches to new physics searches: driven by experimental
      anomalies, driven by model building, or driven by new analysis
      ideas. All of them have their place as long as we keep an open
      mind and see their opportunities as well as their limitations.
    \end{abstract}
  \end{@twocolumnfalse}
]
\thispagestyle{scrheadings}

With the LHC producing its first data relevant for physics beyond the
Standard Model we are looking at exciting times for a theorist
interested in understanding the TeV scale. Let me first review why the
TeV scale is so interesting for particle physicists~\cite{review}:
there exists a whole list of experimental and theoretical shortcomings
of our Standard Model which need to be taken care of by the
ultraviolet completion of our Standard Model which we should really
consider an effective theory.

Experimentally, we do not understand the origin of dark matter;
neither do we understand the source of flavor structures in the quark
or lepton sector, the origin of the matter-antimatter asymmetry in our
universe, the apparent but not quite exact gauge coupling unification,
or any kind of link between our gauge theories with gravity (if that
should exist). Such a list constitutes a good definition of an
effective theory: a theory which has a cutoff separating the
(electroweak) physics we understand and additional observations which
cannot be explained. 

Theoretically, considering the Standard Model an effective theory
bears complications for the Higgs sector. Namely, if there exists a
physical cutoff scale to the Standard Model we should be able to
compute quantum corrections to the Higgs mass in the presence of this
cutoff scale. Such quantum corrections have to be in part absorbed in
a counter term, but given that we construct gauge theories basically
to protect ourselves by {\sl ad-hoc} appearances of unexplained
cancellations we should try to protect the Higgs mass by some kind of
symmetry beyond the Standard Model. This is the motivation of the many
models completing the Standard Model in the ultraviolet. The
corresponding mass range is the scale of new physics which comes in to
protect the Higgs mass. If we want to limit the cancellations between
Standard Model and (the divergence cancelling) new physics
contributions to the Higgs mass the TeV scale turns out to be the key.

This introduction fixes the structure of this talk: aside from general
considerations about models visible with very little luminosity at the
LHC we can follow a theoretically or an experimentally motivated
path. The top forward-backward asymmetry is one example for an
unexplained observation which might or might not require a new physics
explanation. Dark matter is another anomaly, which we are sure exists,
but which might or might not be related to the TeV scale. From a
theory perspective there exist recently considered directions we can
pursue, be it non-minimal supersymmetric models or additional chiral
generations. Finally, from an analysis perspective there are new ideas
appearing which justify a fresh look at the LHC (where the second
example was only partly understood at the time of the talk but is
fully discussed in these proceedings).

The ironic aspect of all of these paths is that we are missing the
obvious one: if the main problem with the Standard Model occurs in the
Higgs sector, why don't we find the Higgs boson first, take our time
to confirm that it is a single fundamental
particle~\cite{sfitter_higgs}, and deal with the theoretical
complications later. This option does not exist because the Higgs
boson only reluctantly couples to protons, so at the LHC we are likely
to only find it after studying the TeV scale for quite a while.

\newpage
\section{First data: supermodels}
\label{sec:supermodels}

In particular analyzing very low luminosities like $10~\text{pb}^{-1}$
at the LHC puts us into an awkward position. Of course we are probing
a new energy scale, and of course we need to analyze data to
understand the way the Standard Model look in brand new detectors, but
there is simply not a lot to discover once we take into account limits
from LEP or from the Tevatron. The fundamental reason is that a 7~TeV
hadron collider does not simply probe all kinds of new physics which
exists up to masses of 3.5~TeV in the case of pair production of new
particles. Instead, most quark and gluon initiated processes happen at
much lower center-of-mass energy and the few events probing larger
energy scales we have to extract statistically. This means that at a
hadron collider luminosity is at least as important as the proton
energies, most notably for gluons where in the interesting energy
range the gluon parton densities scale with the gluon-to-proton
momentum fraction $x$ proportional to $1/x^2$.

In Ref~\cite{supermodels}, which will if nothing else be remembered
for its title, the authors ask the question: what kind of new physics
(supermodels) can we see with at least 10 events in
$10~\text{pb}^{-1}$ without it being ruled out by LEP, flavor physics,
or the Tevatron. Of course, by now we know the answer: nothing was
found. The original argument is nevertheless instructive.\smallskip

\begin{figure}[t]
\includegraphics[width=0.90\hsize]{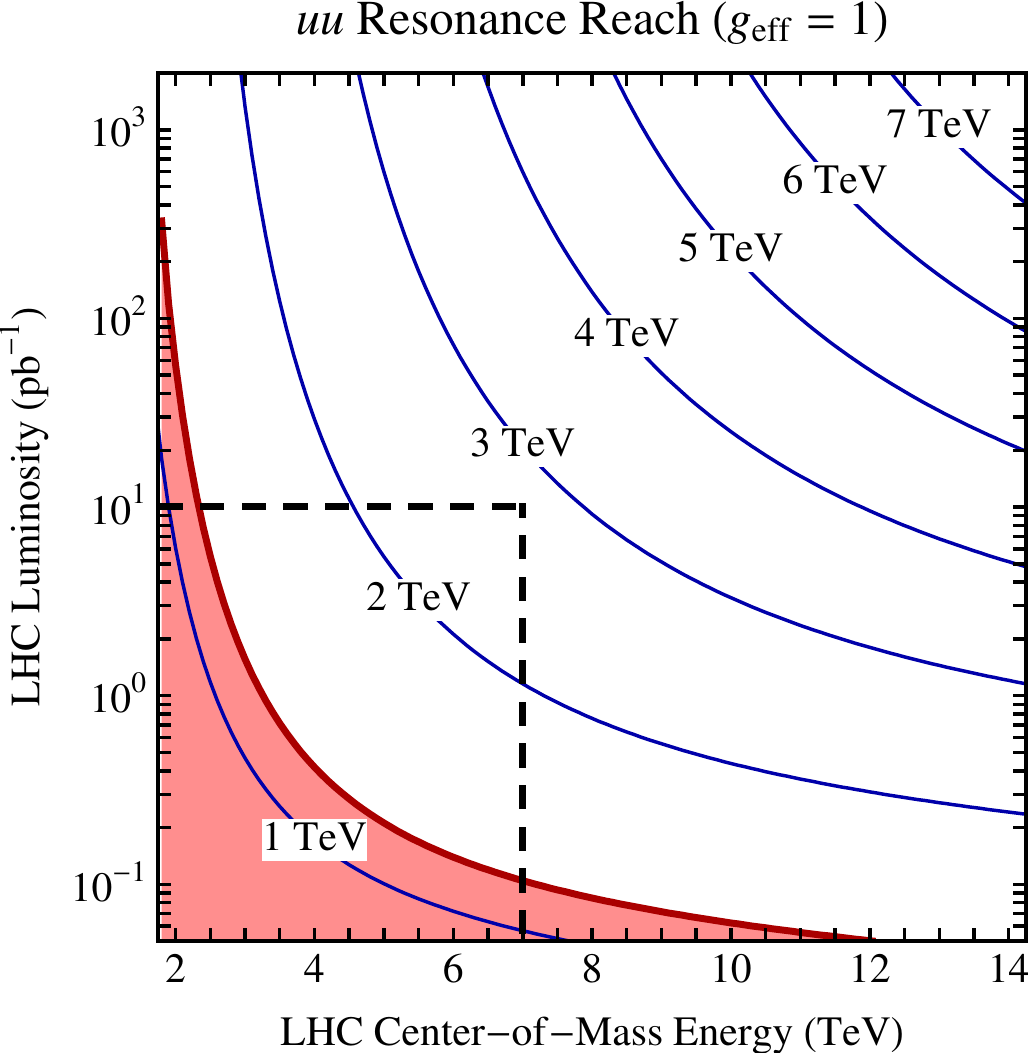} 
\vspace*{-5mm}
\caption{Early LHC reach for in quark-quark scattering. The blue lines
  indicate the mass of the new particle resonantly produces. The red
  shaded region is ruled our by the Tevatron with
  $10~\text{fb}^{-1}$. The dashed lines show the LHC parameters.}
\label{fig:super}
\end{figure}

To begin with, new physics particles can only be produced and decay in
a few relevant topologies. If we are very limited in luminosity and
energy only direct or resonant $s$-channel production is
promising. Supermodels fulfilling the above requirements do not
include pair production even of strongly interacting particles simple
because the LHC cross sections are too small. Including any kind of
branching ratio makes matters worse. This means that for example
searches for leptoquark pair production are not promising for very
early LHC running. For example in gluon fusion resonant production of
$X$ probes the dimension-5 operator
\begin{equation}
\frac{g_s^2}{16 \pi^2 \Lambda} \; X G_{\mu \nu} G^{\mu \nu} \; ,
\end{equation}
which we know from Higgs production. This structure we can write using
the effective coupling $g_\text{eff} = 1/(4 \pi) \, m_X/\Lambda$. In
term of $g_\text{eff}$ we can estimate the reach of different partonic
initial states at the LHC. For a 7~TeV the only promising initial
state to produce new particles predicted by supermodels is the
quark-quark initial state. In Figure~\ref{fig:super} we show how early
LHC analyses can surpass Tevatron analyses in this channel. All other
channels, including the usually most powerful quark-gluon initial
state, are not competitive and have to wait for higher LHC
luminosities.\smallskip

The question remains: what are such supermodels? In resonance
production we usually think of $Z'$ searches first. However, LEP
limits on the mass combined with a leptonic branching ratio solidly
rule out any early LHC discovery before we even start the
analysis. Instead, we can look for di-quark resonances decaying in
easy to observe ways. One of them would be a di-quark $D$ decaying via
a lepto-diquark $L$ as $D \to \ell^- L \to \ell^- \ell^+ +$2 jets. As
strange as this signature looks it is similar to sbottom decays in $R$
parity violating supersymmetry:
\begin{equation} 
\begin{array}{l}
\tilde{b}^c \to b\, \tilde{\chi}_1^0 \\
\phantom{\tilde{b}^c \to b\ } \bentarrow \ell^+\, \tilde{\ell} \\[-2pt]
\phantom{\tilde{b}^c \to b\ \bentarrow \ell\,} \bentarrow \ell^-\, 3j 
\end{array} 
\end{equation}
This specifically chosen fully reconstructable leptonic decay chain
also leads us to a bottom line: new physics at the LHC does not only
need to couple to quarks or gluons to be produced, we also need to see
it in its decay products. This is where backgrounds hurt and where
promising analyses die.

\newpage
\section{Anomaly: top asymmetry}
\label{sec:topfb}

The situation with the observed top forward-backward anomaly at the
Tevatron resembles the situation with the Higgs sector and new physics
at the TeV scale for the LHC. On the one hand, there exists an anomaly
as an experimental puzzle, as presented at this workshop. For a
forward-backward charge asymmetry which in the Standard Model is only
induced at next-to-leading order~\cite{german} the observed value
\begin{equation}
A^\text{exp}_\text{FB} = 0.193 >
A^\text{SM}_\text{FB} = 0.05
\end{equation}
is considerably larger than the expected value. The problem with an
experimental confirmation at the LHC is that there gluon fusion
usually dominates over quark-antiquark scattering which makes it hard
to confirm this Tevatron measurement.\smallskip

So while we would like to learn more about the asymmetry itself we
will be looking for possible new physics scenarios responsible for the
top charge asymmetry. This new physics has fairly specific features
which are not automatically present in our usual new physics scenarios
at the LHC. First, the asymmetry is large, so the responsible new
physics needs to couple to quarks or/and gluons strongly
enough. Second, it needs to generate a charge asymmetry with the
correct sign.

\begin{table}[b]
\renewcommand{\arraystretch}{1.05}
\begin{small} \begin{tabular}{@{}c@{}c@{}c|c@{}c@{}c@{}c|c|c} \hline
\multicolumn{3}{c|}{$SU(3)_1$}&\multicolumn{4}{c|}{$SU(3)_2$}
  & $\Delta A_\text{FB}$ & \\  \hline
&&& $(t,b)_L$&$q_L$&$t_R, b_R$&$q_R$&     $=0$ & coloron \\  
&&$q_R$ & $(t,b)_L$&$q_L$&$t_R, b_R$&  &  $=0$ & \\
&$t_R, b_R$&  & $(t,b)_L$&$q_L$&&$q_R$ &  $=0$ & \\
$q_L$&&  & $(t,b)_L$&&$t_R, b_R$&$q_R$ &  $=0$ & \\
$q_L$&$t_R$, $b_R$ && $(t,b)_L$&&&$q_R$&  $>0$ & candidate \\
$q_L$&& $q_R$ & $(t,b)_L$&&$t_R, b_R$& &  $=0$ & top-color \\
&$t_R, b_R$& $q_R$ & $(t,b)_L$&$q_L $&&&  $<0$ & axigluon \\
$q_L$&$t_R, b_R$&$q_R$ & $(t,b)_L$&&&&    $=0$ & \\ \hline
\end{tabular} \end{small}
\vspace*{-3mm}
\caption{$SU(3)_1 \times SU(3)$ charge assignments in different models
  and their forward-backward asymmetry prediction for top pairs at the
  Tevatron.}
\label{tab:topfb}
\end{table}

The latter turns into a disappointment because there exists a prime
candidate for such an asymmetry, namely an axigluon arising from a
breaking of a $SU(3)_1 \times SU(3)_2$ symmetry into massless QCD and
a remaining $SU(3)$ with a large gauge boson mass $m_C$. To illustrate
the situation, in Tab~\ref{tab:topfb} we show the charge assignments
of the different quark generations and the predicted sign of a
forward-backward asymmetry~\cite{sekhar}. We need $g_1 \ne g_2$ with a
flipped axigluon charge assignment where the left-handed
light-generation is aligned with the right-handed 3rd
generation. Their couplings in terms of the mixing angle have the
structure $g_L^t \sim g_R^q \sim 1 - \cos^2 \theta$.

\begin{figure}[t]
\includegraphics[width=0.85\hsize]{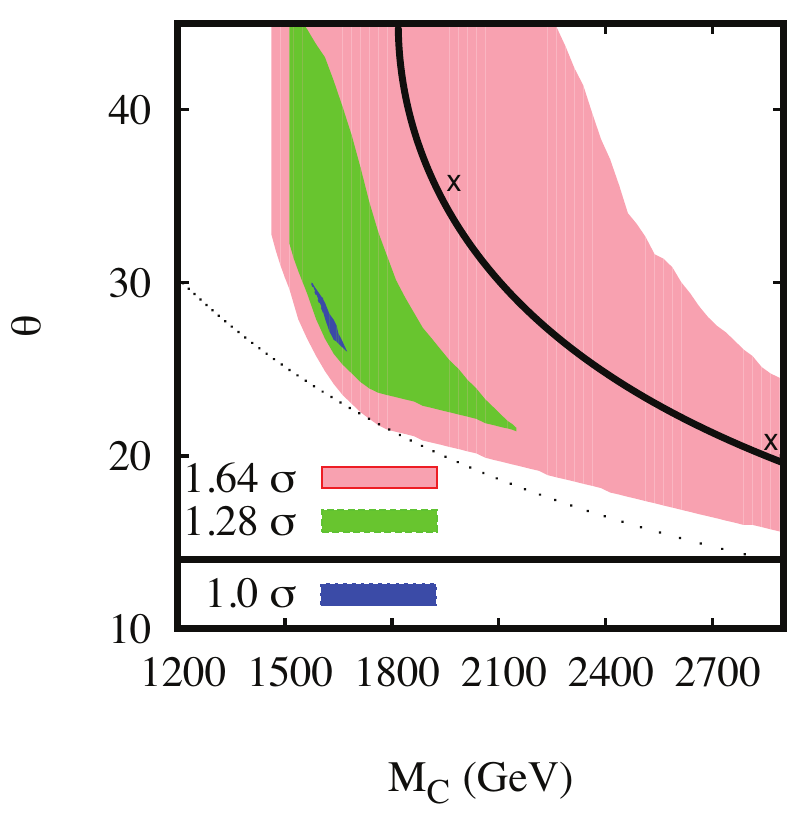} 
\vspace*{-5mm}
\caption{Switched axigluon parameter space. Different levels of
  agreement with data correspond to the different shading. The solid
  black line shows the 95\% CL limit from $B_d$ mixing. The crosses
  correspond to parameter points for which the predicted Tevatron
  anomaly does not exceed 0.05.}
\label{fig:topfb}
\end{figure}

Studying such scenarios requires us to take into account additional
constraints. $B_d$ mixing in certain flavor models requires in terms
of the mass and mixing angle $m_C \sin 2 \theta > 1.8$~TeV, while
electroweak precision data implies $m_C > \cot \theta \times 700$~GeV,
as shown in Fig~\ref{fig:topfb}. Asymmetries resulting from switched
axigluons respecting these constraints do not predict contributions
exceeding $A_\text{FB} \sim 0.04$ and are hence not well suited to
explain the Tevatron anomaly. Different flavor assumptions weaken
these constraints but leave untouched the problem of generating the
large observed asymmetry while staying in agreement with known
constraints~\cite{joanne_tom,uli}.\smallskip

Instead, at the LHC we have to search for models which might explain
this asymmetry in channels which have nothing to do with the
asymmetry. This includes colored particles which violate flavor in $t$
channel exchange. Alternatively, we might see like-sign top pair
production at the LHC.  Luminosities of the order of
$5~\text{fb}^{-1}$ at 7~TeV are already promising for such searches
and might well tell us something about new physics at the TeV scale
long before we get to test if it has anything to do with the Tevatron
anomaly --- or the hierarchy problem, for that matter.

\newpage
\section{Anomaly: dark matter}
\label{sec:dm}

Of all experimental anomalies affecting the Standard Model the
observed dark matter density is certainly the best
established~\cite{dan_review}.  The WIMP miracle implies that a weakly
interacting dark matter particle needs a mass around the TeV scale to
reproduce the observed relic density of cold dark matter. This might
or might not be a coincidence, but for new physics models it is a
strong motivation to introduce a stable new particle. Its stability
can be ensured by a $Z_2$ symmetry which as a side effect can protect
the model from electroweak precision data.\smallskip

\begin{table}[b]
\renewcommand{\arraystretch}{1.05}
\begin{center} \begin{small} \begin{tabular}{c|c|r}
\hline
       & operator & coefficient   \\
\hline
D1 & $\bar{\chi}\chi\bar{q} q$ & $m_q/\Lambda^3$   \\
D2 & $\bar{\chi}\gamma^5\chi\bar{q} q$ & $im_q/\Lambda^3$    \\
D3 & $\bar{\chi}\chi\bar{q}\gamma^5 q$ & $im_q/\Lambda^3$    \\
D4 & $\bar{\chi}\gamma^5\chi\bar{q}\gamma^5 q$ & $m_q/\Lambda^3$   \\
D5 & $\bar{\chi}\gamma^{\mu}\chi\bar{q}\gamma_{\mu} q$ & $1/\Lambda^2$   \\
D6 & $\bar{\chi}\gamma^{\mu}\gamma^5\chi\bar{q}\gamma_{\mu} q$ & $1/\Lambda^2$    \\
D7 & $\bar{\chi}\gamma^{\mu}\chi\bar{q}\gamma_{\mu}\gamma^5 q$ & $1/\Lambda^2$   \\
D8 & $\bar{\chi}\gamma^{\mu}\gamma^5\chi\bar{q}\gamma_{\mu}\gamma^5 q$ & $1/\Lambda^2$   \\
D9 & $\bar{\chi}\sigma^{\mu\nu}\chi\bar{q}\sigma_{\mu\nu} q$ & $1/\Lambda^2$   \\
D10 & $\bar{\chi}\sigma_{\mu\nu}\gamma^5\chi\bar{q}\sigma_{\alpha\beta}q$ & $i/\Lambda^2$  \\
D11 & $\bar{\chi}\chi G_{\mu\nu}G^{\mu\nu}$ & $\alpha_s/4\Lambda^3$   \\
D12 & $\bar{\chi}\gamma^5\chi G_{\mu\nu}G^{\mu\nu}$ & $i\alpha_s/4\Lambda^3$   \\
D13 & $\bar{\chi}\chi G_{\mu\nu}\tilde{G}^{\mu\nu}$ & $i\alpha_s/4\Lambda^3$   \\
D14 & $\bar{\chi}\gamma^5\chi G_{\mu\nu}\tilde{G}^{\mu\nu}$  & $\alpha_s/4\Lambda^3$ \\
\hline
\end{tabular} \end{small} \end{center}
\vspace*{-6mm}
\caption{Operators coupling a Dirac fermion dark matter candidate to
  Standard Model particles and their scaling in terms of an effective
  theory.}
\label{tab:dm}
\end{table}

In the popular minimal supersymmetric Standard Model, the dark matter
candidate is a Majorana fermion with only weak charge. The scalar
neutrino could serve the same purpose, but comparing its annihilation
rate to the observed relic density with direct detection rules is
out. In its short life time the PAMELA anomaly motivated studies of a
Dirac gaugino which would naturally prefer to annihilate to leptons
and not to quarks. Finally, in models with universal extra dimensions
the dark matter candidate is the Kaluza-Klein partner of the neutral
weak gauge bosons. The lesson to learn from these exercises in model
building is that even if we assume that our dark matter agent is a
WIMP, we should not only look for a Majorana fermion.\smallskip

An interesting observation is that at the amplitude level we can link
the scattering of a WIMP off a nucleon in direct detection with WIMP
pair production at the LHC
\begin{equation}
q \chi \to q \chi \qquad \Leftrightarrow \qquad 
q \bar{q} \to \chi \chi^* \; ,
\label{eq:dm}
\end{equation}
where the star generically implies the anti-WIMP. Such a process can
for example be mediated by an $s$-channel $Z$ or Higgs boson or a
$t$-channel partner of the quark, if such a particle should exist. For
supersymmetry similar links have been successfully established between
direct detection experiments and pseudoscalar Higgs production at
colliders, while any correlation is absent between dark matter
observables and neutralino-chargino pair production decaying to
tri-leptons~\cite{with_dan}.

Avoiding specific models, to a given dimension we can list all
operators which can mediate these processes for a Dirac fermion dark
matter candidate in Tab~\ref{tab:dm}~\cite{tim}. Similar lists we can
produce to a Majorana fermion or real or complex scalars. Note that
this list of operators includes those for example mediated by
$t$-channel squarks in the case of supersymmetry, but in the limit of
decoupled squarks. 

\begin{figure}[t]
\includegraphics[width=0.99\hsize]{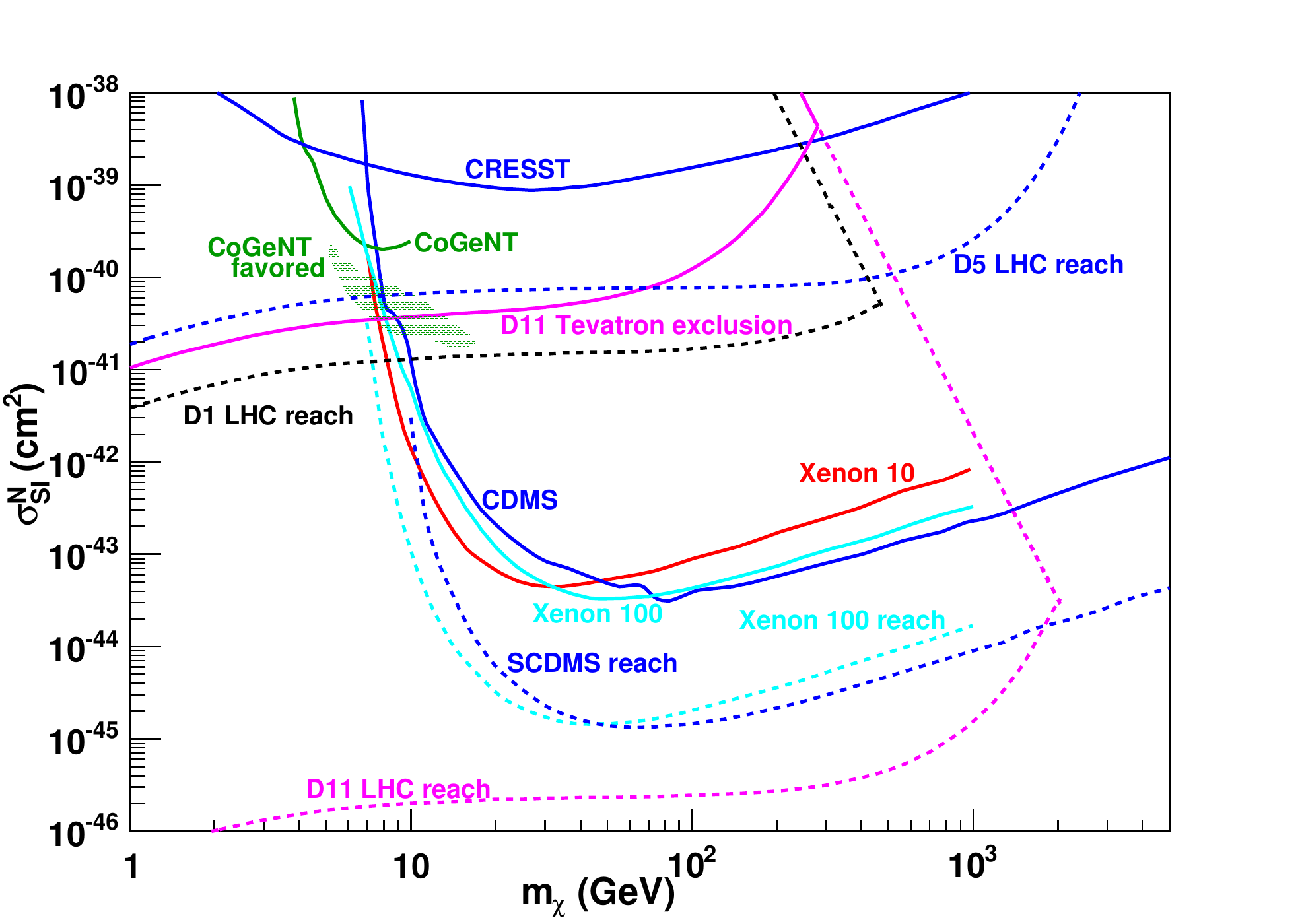} 
\vspace*{-5mm}
\caption{Limits on Dirac fermion dark matter from spin-independent
  direct detection searches and (projected) from LHC, based on the
  operators in Tab~\ref{tab:dm}.}
\label{fig:dm}
\end{figure}

This list of operators we can analyze for the direct detection and for
the collider direction of the process shown in Eq.(\ref{eq:dm}). In
Fig~\ref{fig:dm} we see how for example spin-independent direct
detection and LHC reaches compare~\cite{tim}.  The only caveat of such
a comparison is that it might drastically underestimate the LHC reach,
since at hadron colliders we can produce any light enough strongly
interacting particle directly and should not integrate it out in an
effective theory. Squark pair production decaying to neutralinos in
supersymmetry is a good example for this situation, with
$\sigma(\tilde{q} \tilde{q} \to \chi\chi + X) \gg \sigma(\chi \chi)$.

\newpage
\section{Models: Dirac gauginos}
\label{sec:mrssm}

A major problem with the minimal supersymmetric Standard Model are the
many constraints from the flavor sector, including electric dipole
moments. While a complete absence of any signal for new physics can be
considered a sign for particular symmetry structures at and beyond the
TeV scale they are also worrisome. The question arises, if it is
possible to alleviate the pressure the non-observation of any
higher-dimensional flavor operators puts on the MSSM.

\begin{figure}[b]
\begin{center}
\includegraphics[width=0.90\hsize]{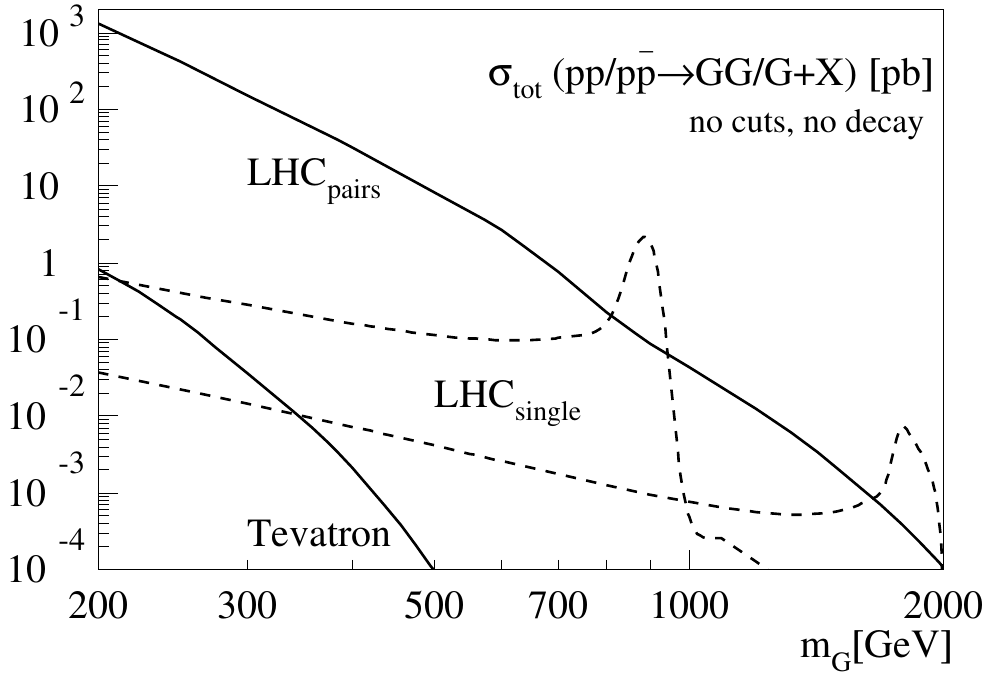} 
\end{center}
\vspace*{-8mm}
\caption{Cross sections for sgluons at the Tevatron and a 14~TeV
  LHC. For the LHC we show pair (solid) and single production
  (dashed). The two curves for single sgluon production assume a
  gluino mass of 1~TeV and squark masses of 500~GeV (upper) and
  1~TeV (lower).}
\label{fig:sgluon1}
\end{figure}

One way to do this is to promote the $R$ parity we need to prevent
proton decay and to stabilize a dark matter candidate to a continuous
global symmetry. If we also avoid spontaneous supersymmetry breaking
such a symmetry forbids a fair fraction of soft breaking terms,
including Majorana masses, tri-scalar interactions, or the $\mu$
term. The absence of these terms gets rid of many dimension-5 flavor
violating operators, but it leaves open how to give mass for example
to the gluino~\cite{mrssm}. With two degrees of freedom mirroring the
two gluon polarizations the gluino stays massless unless we identity
two additional degrees of freedom to generate a Dirac mass. The
solution is to introduce a complex scalar sgluon~\cite{sgluon} with a
Standard Model $R$ charge. This sgluon we can also interpret as a
result of an extended $N=2$ supersymmetry which allows us to
consistently extrapolate between Majorana and Dirac gluino masses in a
phenomenological analysis~\cite{n2susy}.  The same feature of course
appears for the weak gauginos, but its experimental signatures are
much less generic.

From regular minimal supersymmetry we know that the SUSY-QCD sector is
especially predictive. The same is the case for the gluino-sgluon
sector: the sgluon's adjoint color charge fixes the
gluon-sgluon-sgluon coupling, supersymmetry predicts a strong
sgluon-gluino-gluino coupling, and $D$ terms determine the strong
squark-squark-sgluon-sgluon coupling. This fully determines the pair
production cross section at hadron colliders.

\begin{figure}[t]
\begin{center}
\includegraphics[width=0.70\hsize]{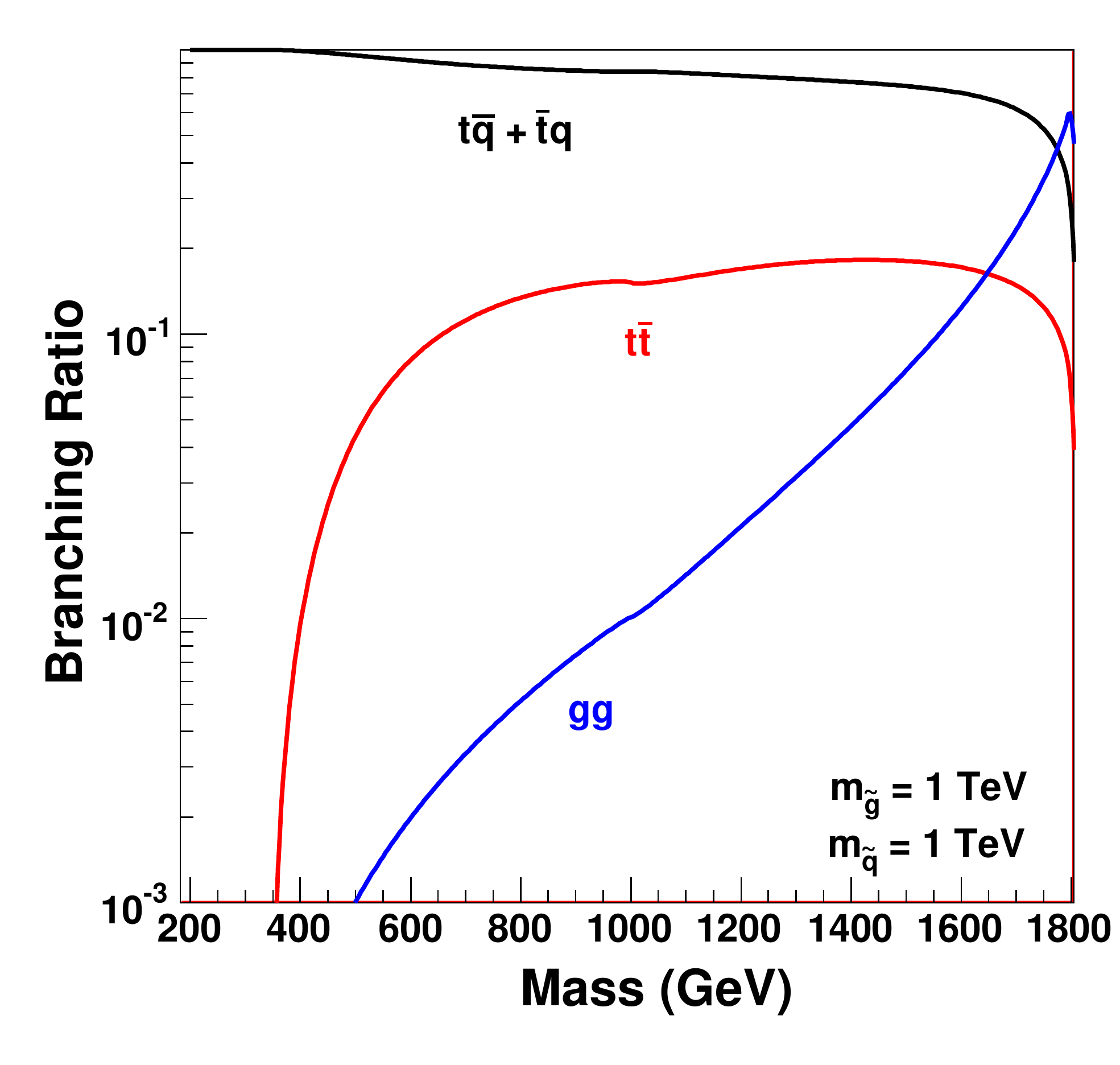} 
\end{center}
\vspace*{-8mm}
\caption{Sgluon branching ratios for different left-handed squark and
  gluino masses.  Right-handed squarks are set to $90\%$ of the
  left-handed squark masses. We assume maximal up-squark mixing.}
\label{fig:sgluon2}
\end{figure}

Couplings involving a single sgluon, relevant for single production or
sgluon decays, are generated at the one-loop level. As we can see
in Fig~\ref{fig:sgluon1} the pair production rate exceeds single
production unless on-shell effects inside the loop enhance the
latter. An interesting structure appears in the quark-quark-sgluon
coupling which is proportional to the quark mass and due to the
weakened flavor constraints does not have to be flavor diagonal. In
Fig~\ref{fig:sgluon2} we see how only fairly heavy sgluons will not
decay to $q\bar{t} + t\bar{q}$ but to a pair of gluon jets. At the LHC
we would observe two like-sign tops, each together with a
hard jet reconstructing the sgluon.

An interesting alternative search is based on ignoring the motivation
of the $R$-symmetric MSSM and instead look for generic multi-jet final
states without leptons or missing energy or any other distinctive
feature but a set of mass constraints. Any kind of color octet like
axigluons (as discussed in Sec~\ref{sec:topfb}), colorons, or heavy
gluon partner predicts such signatures. In the case of colorons it has
been nicely shown how given enough hard jets and mass constraints we
can not only identify a new physics signature but also determine the
masses of the particle involved~\cite{coloron}.

\newpage
\section{Models: four generations}
\label{sec:g4}

Searches for four generations are an obvious task for the LHC, given
that the number of chiral fermion generations is not linked to any
property of the Standard Model as a field theory. There exists a
multitude of motivations for four generations, weakly as well as
strongly interacting, with and without supersymmetry, etc.

Arguments used against four generations include 
\begin{itemize}
\item[--] mass degenerate heavy up and down quarks are forbidden by
  electroweak precision data --- the top and bottom mass in the
  Standard Model are not degenerate, either, and a 10\% mass splitting
  cures this problem~\cite{g4us}.
\item[--] the theory might become strongly interacting at large
  energies --- this might either lead to electroweak symmetry
  breaking~\cite{holdom} or be absent due to the fixed point structure
  of the Yukawas and the Higgs self
  coupling~\cite{christof1}.
\item[--] a generic unhappiness with heavy fourth generation neutrinos ---
  which can be avoided altogether for some more exotic unified group
  representations~\cite{christof2}.
\item[--] or simply the feeling that `there should not be a fourth
  generation'~\cite{peccei}.
\end{itemize}
Obviously, there exist at least as good arguments against
supersymmetry, extra dimensions, or strongly interacting Higgs
models. An easy way to cure the problem with electroweak precision
data is shown in Tab~\ref{tab:g4}: with the correct ordering and a
small splitting between the heavy $u_4$ and $d_4$ quarks we move along
the main axis of the $S$ vs $T$ ellipse. As a side effect any Higgs
mass is allowed in these models.\smallskip

\begin{table}[b]
\renewcommand{\arraystretch}{1.05}
\begin{center} \begin{small} \begin{tabular}{ccc|cc}
$m_{u_4}$ & $m_{d_4}$ & $m_H$ & $\Delta S_\text{tot}$ & $\Delta T_\text{tot}$
\\ \hline
$310$ & $260$ & $115$ & $0.15$ & $0.19$ \\
$320$ & $260$ & $200$ & $0.19$ & $0.20$ \\
$330$ & $260$ & $300$ & $0.21$ & $0.22$ \\ \hline
$400$ & $350$ & $115$ & $0.15$ & $0.19$ \\
$400$ & $340$ & $200$ & $0.19$ & $0.20$ \\
$400$ & $325$ & $300$ & $0.21$ & $0.25$ \\
\end{tabular} \end{small} \end{center}
\vspace*{-6mm}
\caption{$\Delta S$ and $\Delta T$ from a fourth generation. The
  lepton masses are $m_{\nu_4} = 100$~GeV and $m_{\ell_4} = 155$~GeV,
  giving $\Delta S_{\nu \ell} = 0.00$ and $\Delta T_{\nu \ell} =
  0.05$.  All points are within the 68\% CL contour defined by the
  LEP~EWWG.}
\label{tab:g4}
\end{table}

A likely effect of electroweak precision data, but not
unsurmountable~\cite{alex} is that the heavy $u_4$ should decay to the
$d_4$ and not vice versa. This could give rise to decay cascades of
the kind
\begin{equation}
u_4 
\to (W d_4)
\to (W W t) 
\to (W W W b) \; ,
\end{equation}
observable for example as many like-sign leptons.\smallskip

\begin{figure}[t]
\includegraphics[width=0.99\hsize]{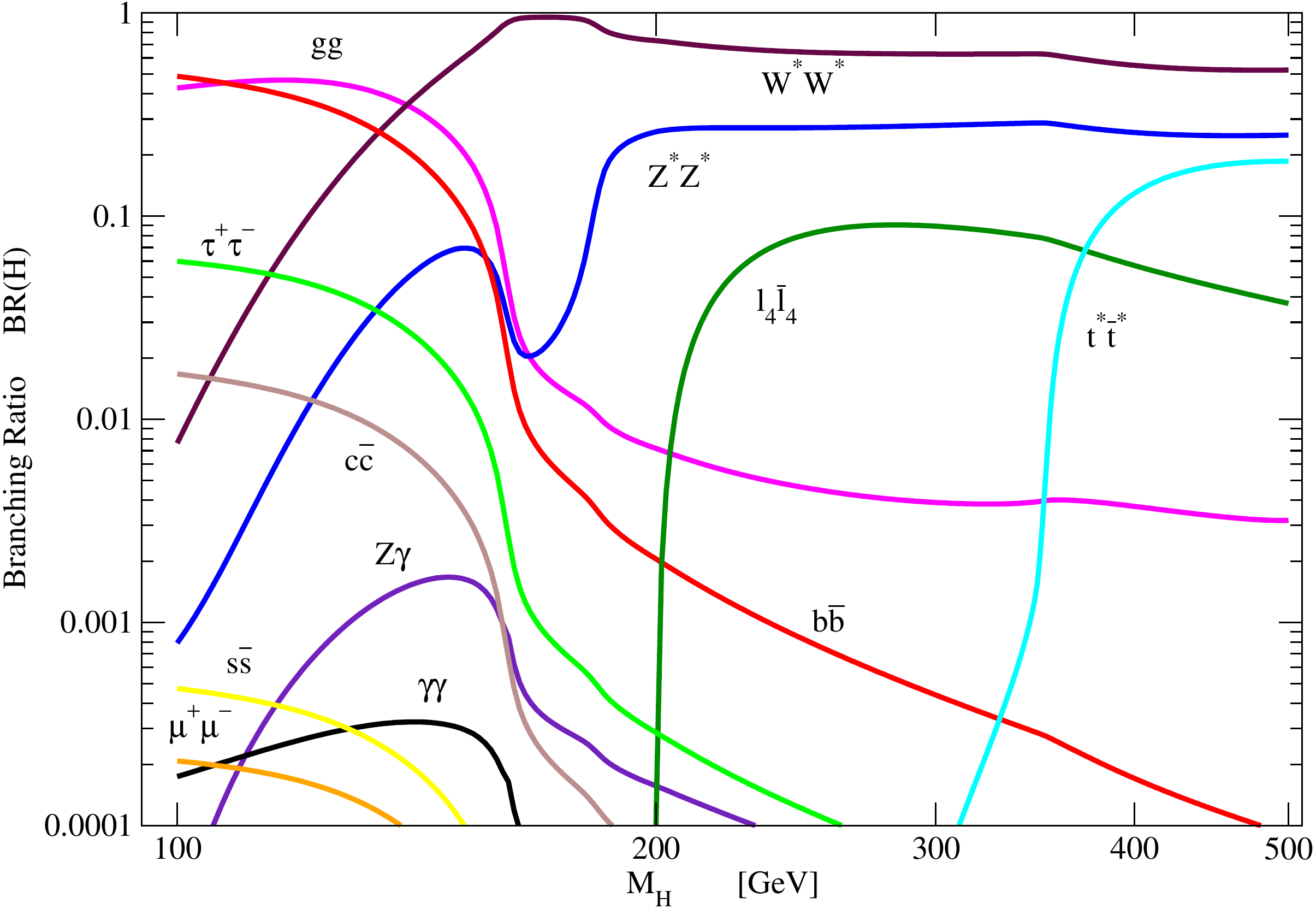} 
\vspace*{-5mm} 
\caption{Higgs branching ratios based on the mass spectrum shown in
  the second line in Tab~\ref{tab:g4}. The numbers are obtained with a
  modified version of {\sc Hdecay}~\cite{hdecay}.}
\label{fig:g4}
\end{figure}

Chiral fermions have the particular property that they do not decouple
from a theory. For example, in the dimension-5 gluon-gluon-Higgs
coupling the Yukawa coupling in the numerator cancels the decoupling
heavy mass in the denominator. For any quark the coupling is only
suppressed by the Higgs VEV and approaches a constant value for large
quark masses and Yukawas. A fourth quark generation roughly triples
the effective Higgs coupling to gluons or increases the leading LHC
cross section for Higgs production by a factor 9. For small Higgs
masses such an enhancement is ruled out by the Tevatron, but this
constraint vanishes if the Higgs becomes heavier according to
Tab~\ref{tab:g4}. For small Higgs masses the tree-level decays to tau
leptons and $b$ quarks are surpassed by the similarly enhanced loop
induced decay to two gluons, which will essentially be invisible at
the LHC. For large Higgs masses the decays to $W$ and $Z$ bosons which
in the heavy Higgs limit scale like 2:1 are unaffected.

When we add supersymmetry to models with a fourth generation we see
two interesting effects: first, electroweak baryogenesis can be
revived with a fairly small mass splitting between fourth
generation quarks and squarks. Second, we can generate the entire
light CP-even Higgs mass via loop effects and avoid the little
hierarchy problem altogether. We might even see a supersymmetric light
CP-even Higgs decaying to $W$ pairs~\cite{4mssm}.

\newpage
\section{Analysis: boosted tops}
\label{sec:boosted}

Looking for new physics using boosted heavy Standard Model particles
decaying to quarks or gluons ($W,H,t$) has long been recognized as a
motivation to study not only the outcome of a jet algorithm, but also
the clustering history~\cite{mike}. Recently, this strategy has
revived a major search channels for a light Higgs at the LHC, namely
$WH/ZH$ production with $H \to b\bar{b}$~\cite{gavin}. Already from
the first attempts to use fat jets from massive particles~\cite{mike}
we know that hadronic top decays are promising candidates to solve
combinatorial issues as well as improve the mass resolution of the
particles decaying to tops.

\begin{figure}[b]
\begin{center}
\vspace*{-8mm}
\includegraphics[width=0.70\hsize]{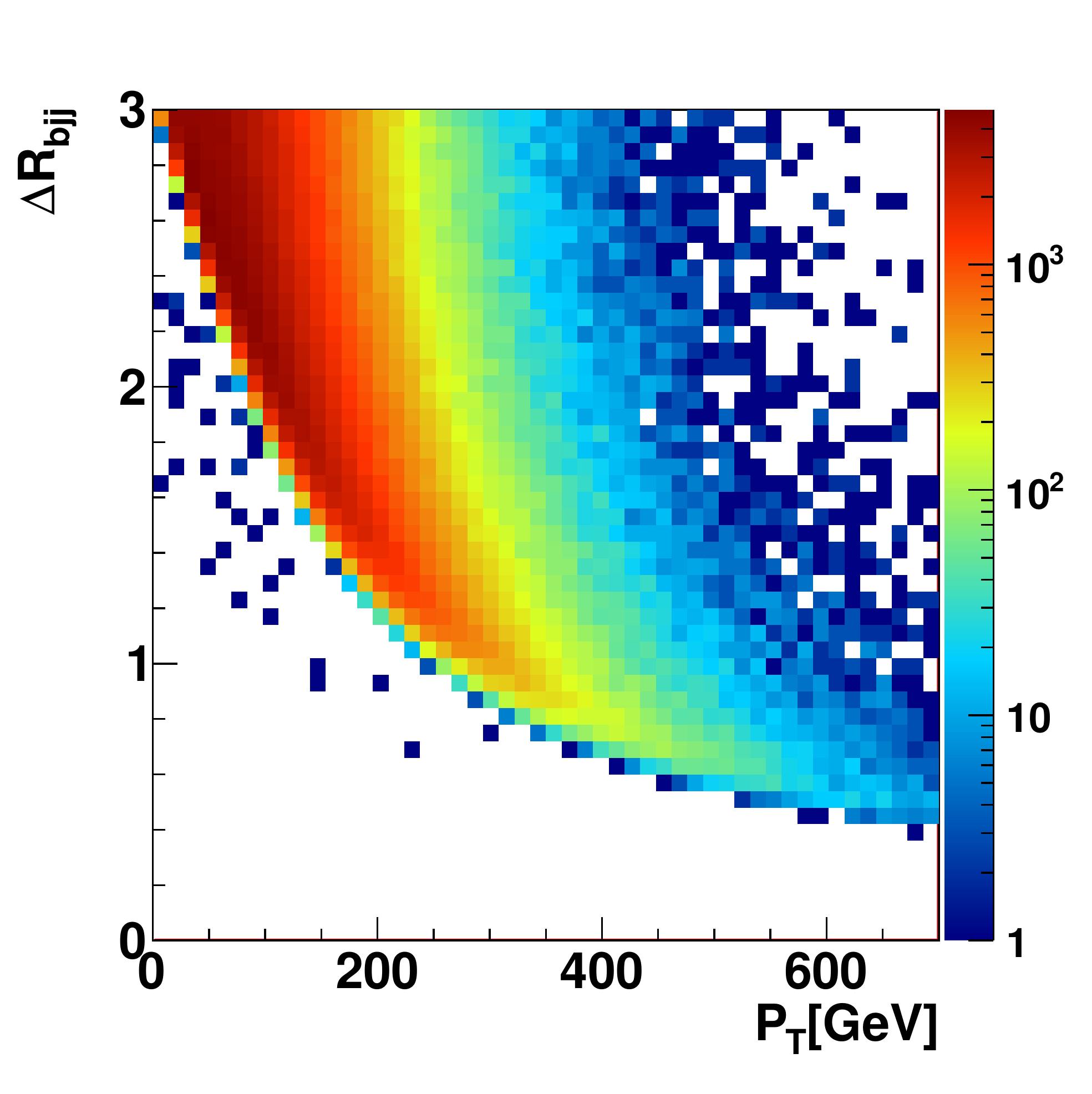} 
\end{center}
\vspace*{-9mm}
\caption{Top pair events at the LHC as a function of the transverse
  momentum and an approximate $R$ size needed to cover all decay
  products.}
\label{fig:boosted1}
\end{figure}

One of the standard top taggers inspired by the $H \to b\bar{b}$
analysis~\cite{gavin} is the Johns Hopkins tagger, optimized for heavy
resonances decaying to very strongly boosted top
pairs~\cite{hopkins}. The problem of any such specialized taggers is
that they are unlikely to be tested in the Standard Model $t\bar{t}$
sample and that their range of applications is limited. In
Fig~\ref{fig:boosted1} we show the correlation of kinematic parameters
we expect for more or less boosted top decay products~\cite{heptop}. A
maximum usable jet size around $R=1.5$ translates into a transverse
momentum $p_{T,t} > 200$~GeV. The challenge is to construct a top
tagger which works down to this kind of transverse momenta.

\begin{figure}[t]
\begin{center}
\includegraphics[width=0.78\hsize]{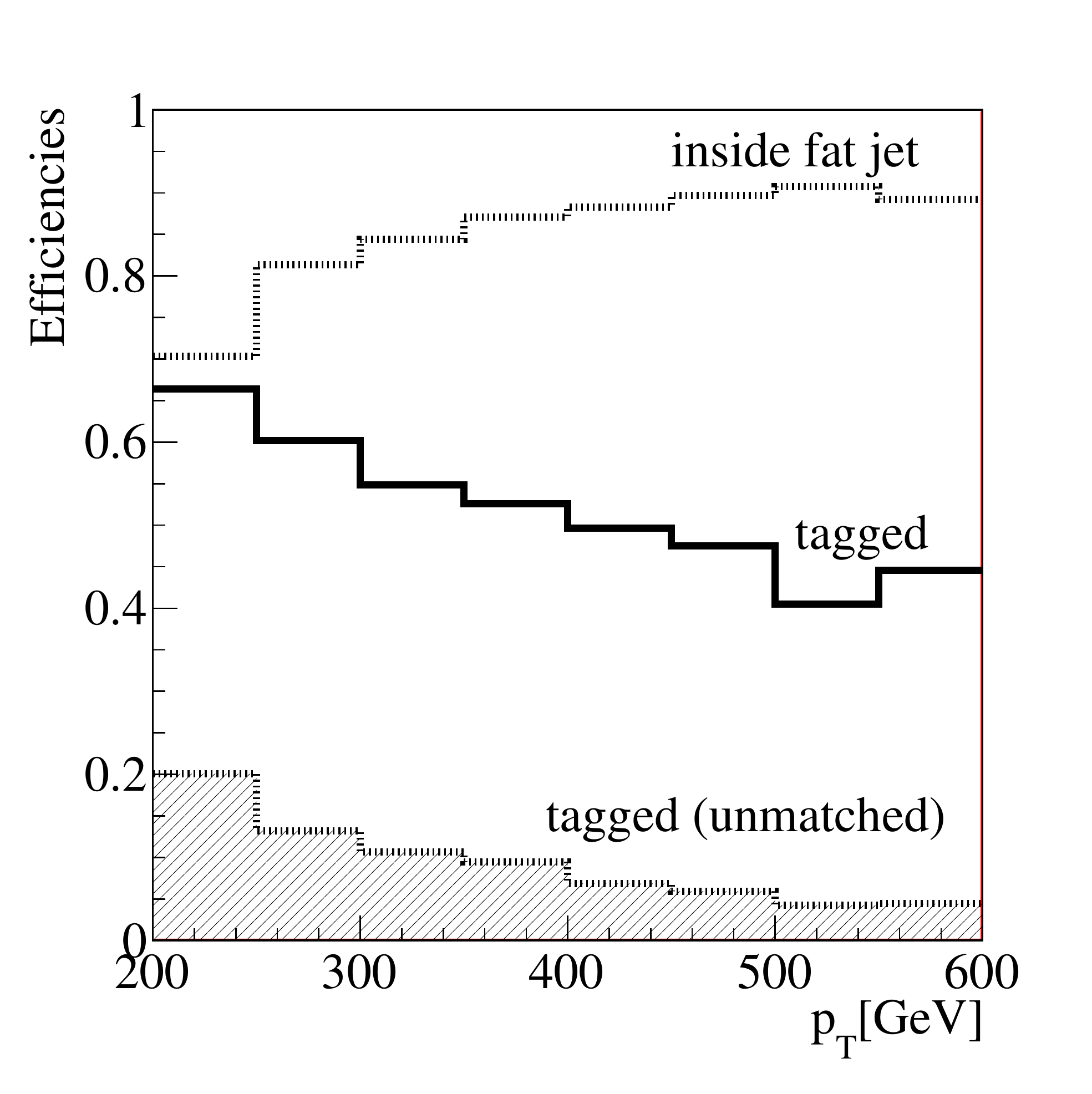} 
\end{center}
\vspace*{-11mm}
\caption{{\sc HEPTopTagger} efficiencies evaluated on $t\bar{t}$
  events normalized to all tops covered by a jet size $R=1.5$. The
  unmatched fraction indicated events where our naive geometric
  mapping of tagged tops to Monte Carlo truth is not unique.}
\label{fig:boosted2}
\end{figure}

The publicly available {\sc HEPTopTagger}~\cite{heptop} has been shown
to resolve the combinatorial and background issues in $t\bar{t}H$
searches, but the required integrated luminosities around
$100~\text{fb}^{-1}$ would benefit from further experimental
studies. The tagging efficiencies we show in
Fig~\ref{fig:boosted2}. Typical mis-tagging rates on QCD or $W$+jet
events range around a few per-cent. A detailed ATLAS study is on the
way.\smallskip

As mentioned in the Introduction, we might well observe for
example a top partner helping to stabilize the Higgs mass before we
see the Higgs boson itself. Searching for top partners decaying
flavor-diagonally to a top quark and a weakly interacting dark matter
agent (as discussed in Sec~\ref{sec:dm}) is one of the best motivated
dedicated searches at the LHC.

Without the help of a top tagger searching for top partners decaying
to hadronic or semi-leptonic top pairs is not promising due to
overwhelming systematic uncertainties and a limited
signal-to-branching ratio $S/B \sim 1/10$~\cite{heptop}. In contrast,
a successful hadronic analysis based on tagged tops is essentially the
same as a slepton or sbottom pair search where we reconstruct the
decay lepton's or bottom's 4-momentum and apply an $M_{T,2}$ cut.  It
allows us to extract the stop signal for stop masses from
$350-650$~GeV at a 14~TeV LHC with luminosities around
$10~\text{fb}^{-1}$. Moreover, the clearly visible endpoint of the
$M_{T,2}$ distribution determines the stop mass given the mass of the
dark matter candidate. Because we can fully rely on the reconstructed
top 4-momentum we do not make use of angular correlations, making
such an analysis easily generalizable.

An obvious question is if we can extract boosted semileptonic stop
pairs based on one hadronic and one leptonic tag. It turns out
feasible with luminosities comparable to the hadronic mode, but with a
less impressive leptonic 4-momentum
reconstruction~\cite{leptontagger}.

\newpage
\section{Analysis: inclusive jet searches}
\label{sec:inc}

Recently, the first LHC results on searches for new physics ---
specifically for supersymmetry --- in jets plus missing energy, plus
zero leptons~\cite{lhc_susy}. The questions, following for example the
first CMS paper, is how general we can keep these searches and to what
degree we need to apply specific background rejection cuts to improve
the results of counting experiments.

\begin{figure}[b]
\vspace*{-5mm}
\begin{center}
\includegraphics[width=0.99\hsize]{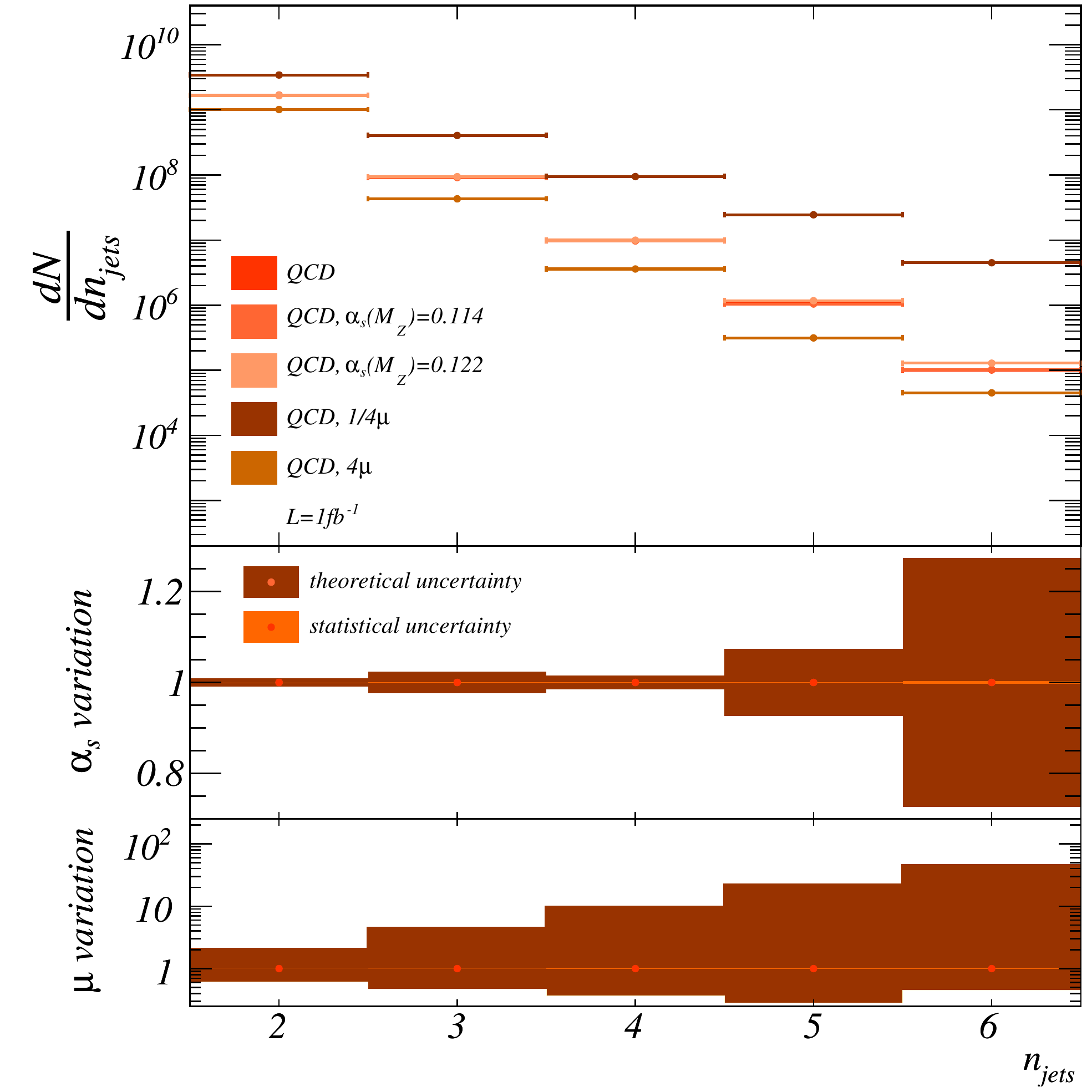} 
\end{center}
\vspace*{-8mm}
\caption{Exclusive $d\sigma/d\nj$ distribution for QCD jets.  The
  second panel shows the parametric uncertainty due to
  $\alpha_s(m_Z)$.  The third panel shows the consistent scale factor
  treatment which can be experimentally determined.}
\label{fig:incl1}
\end{figure}

Ideally, we would only apply a minimal set of cuts to reduce the pure
QCD and $W/Z$+jets backgrounds to a manageable level with respect to
systematic uncertainties, namely 
$\met > 100$~GeV
and no hard lepton.
%
%
Two key observables when looking at QCD final states are the number of
jets $\nj$ and the effective mass $\meff$ in its most inclusive
definition $\meff = \met + \sum_\text{all jets} p_{Tj}$. Both
observables we compute taking into account all jets which fulfill
$|y_j| < 4.5$ and $p_{T,j} > 50$~GeV. The problem with the $\nj$ and
$\meff$ distributions is that they are theoretically not well
studied. Experimentally, we know that the inclusive $\nj$
distributions shows the so-called staircase scaling~\cite{staircase}
\begin{equation}
\frac{\sigma_{n+1}}{\sigma_n} 
= R_{(n+1)/n} 
= \text{const} \; ,
\label{eq:def_r}
\end{equation}
which holds equivalently for inclusive and exclusive jet
rates~\cite{autofocus}. From a statistics point of view, we clearly
prefer the exclusive $\nj$ distribution where each event is only
assigned to one bin.

In Fig~\ref{fig:incl1} we show how this exclusive scaling can be
simulated using {\sc Ckkw} merging as implemented in {\sc
  Sherpa}~\cite{ckkw}.  What is crucial to use this scaling as
background estimates in new physics searches is to understand the
uncertainties. First, we estimate the theory uncertainty due to a
consistent variation of $\alpha_s(M_Z)$ in the matrix elements, the
parton shower, and the parton densities. This error bar is manageable
and might be comparable to the experimental systematics or to
statistical errors after the basic background rejection cuts 
mentioned above.

The bottom panel shows the naive error estimate from varying all
factorization and renormalization scales around a central value
$\mu/\mu_0 = 1/4 - 4$. This scale factor cannot be derived from first
principles. However, we can estimate it from data and find that in
{\sc Sherpa} it should essentially be unity.  Determining this scale
factor and fixing the normalization of the two-jet rate from
experiment allows us to within a well defined theory uncertainty
predict the $\nj$ and $\meff$ distributions for QCD and $W/Z$+jets
backgrounds. Of course, for the new physics channels we need to rely
on simulations for both
distributions~\cite{review,sgluon,autofocus}.

\begin{figure}[t]
\begin{center}
\includegraphics[width=0.70\hsize]{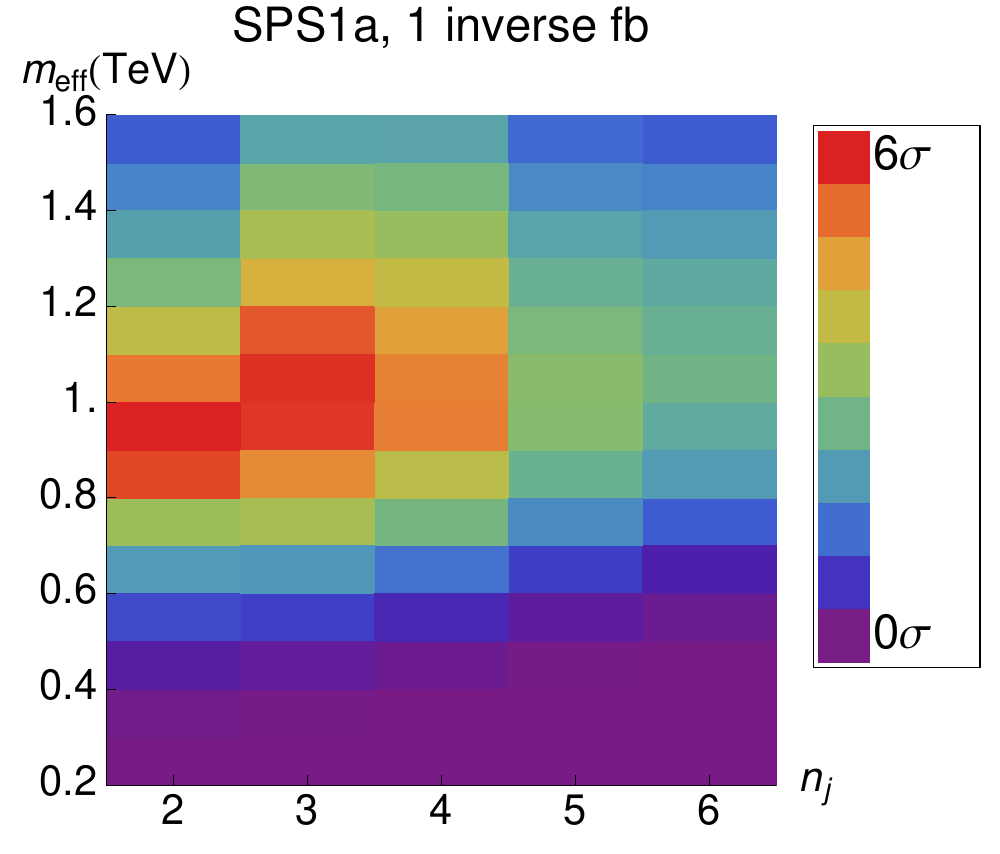} 
\end{center}
\vspace*{-8mm}
\caption{$(\nj,\meff)$ plane for the supersymmetric SPS1a signal
  showing the individual log-likelihoods in each bin.}
\label{fig:incl2}
\end{figure}

From a theory point of view, these two distributions are particularly
interesting because $\meff$ is correlated with the mass of the new
strongly interacting particles and $\nj$ is largely determined by
their color charge. In Fig~\ref{fig:incl2} we shows this
two-dimensional correlation which we can understand in terms of the
different squark and gluino production channels.

\newpage
\section{Exciting times ahead}

The way I organized this review should be considered a discussion of
recent papers I personally found interesting right before I presented
them at Physics in Collisions 2010. It is not meant to be an
overview of recent developments or a thorough review of classical new
physics at the LHC. For the latter, please look at our comprehensive
review article Ref~\cite{review}. Along the same lines, I did not make
any attempt to cover the literature on the seven topics I discussed in
the individual sections. An appropriate coverage of the relevant
publications should be found in the individual papers or newer
publications which have appeared since I gave this talk in the Fall of
2010.\smallskip

What I wanted to illustrate are four different ways to approach new
physics searches at the LHC, in particular in the early stage of LHC
data analysis. Lacking better guidance we can follow a very pragmatic
approach asking what we can actually see given a certain experimental
performance. Second, we can follow more or less well established
experimental anomalies like the top forward-backward asymmetry, weakly
interacting dark matter, or any other anomaly to our liking. Third, we
can let ourselves be inspired by new developments or frequent
recurrences in TeV-scale model building.  Finally, we can get most
excited about progress in LHC analysis techniques, like for example
fat jets, and new opportunities they give us. All these approaches are
equally well motivated and honorable, given that we might well need to
look for the solution of the TeV-scale puzzle before getting to look
at the Higgs sector or the structure of electroweak symmetry breaking
in any detail.\smallskip

What is neither well motivated nor honorable is to let our individual
perspective bias our view of new physics searches: we simply do not
know what to look for at the LHC.  Whatever we might find will at
least prove most of us, possibly all of us wrong. This makes it
crucial to set up and interpret searches in the most general framework
we can. While there does not exist any such thing as a feasible
general search for physics beyond the Standard Model at the LHC, for
example looking for the supersymmetric parameter point SPS1a' at the
LHC makes no sense. Structures have to be inferred from observation
not included in the analysis. A particularly bad example is the recent
CMS analysis~\cite{lhc_susy} which presents the results of the first
supersymmetry search in terms of $m_0$ and $m_{1/2}$. Because there
does not exist a mapping from the $m_0$ vs $m_{1/2}$ to the physical
squark and gluino mass plane the presentation of the results
deliberately excludes scenarios in which the gluino is significantly
heavier than the squark. What the authors seem to not be aware of is
that topologically an ellipse (constant squark mass) and a straight
line (constant gluino mass) do not have to meet.\smallskip

Coming back, the wide open field of new physics is, in my opinion, the
most exciting aspect of LHC searches. Because ATLAS and CMS are
multi-purpose experiments we are not waiting for one specific
observable, like for example the anomalous magnetic moment of the
muon, to either agree or disagree with the Standard Model
predictions. Developments in model building and phenomenology have
shown that new physics at the TeV scale will very generically be
discovered at the LHC, so we can freely choose between the four
sources of inspiration listed above, or anything else.\smallskip

I would like to end this proceedings article with a quote by Uli Baur,
who passed away not only much too early in his life but also at the
most annoying time in his personal physics agenda. Even though Uli
himself will not see the physics discoveries he was looking forward to
for decades I remember his motivation for new physics searches: We
will always discover new physics when we look at much higher energy
scales.

\bigskip
\bigskip
\bigskip
\bigskip
\bigskip

{\footnotesize I would like to thank many colleagues for their insightful
  comments, including Uli Haisch and Susanne Westhoff on flavor
  constraints on axigluons and Graham Kribs and Tim Tait on almost all
  topics discussed in this talk.}

\newpage


\end{document}